\documentclass[aps,prd,reprint,onecolumn,notitlepage,superscriptaddress]{revtex4-1}

\usepackage{lipsum}
\usepackage{amsmath}
\usepackage{graphicx}
\usepackage{color}
\usepackage{tikz}
\usepackage{verbatim}

\usepackage{amssymb}
\usepackage{bm}

\newcommand{\rom}[1]{\textup{\uppercase\expandafter{\romannumeral#1}}}

\begin{document}

\title{Blackhole in Nonlocal Gravity: Comparing Metric from Newmann-Janis Algorithm with Slowly Rotating Solution}
\author{Utkarsh Kumar\footnote{email:kumaru@ariel.ac.il},
       }
\affiliation{Department of Physics, Ariel University, Ariel 40700, Israel}

\author{Sukanta Panda\footnote{email:sukanta@iiserb.ac.in},
        Avani Patel\footnote{email: avani@iiserb.ac.in}}
        \affiliation{ Indian Institute of Science Education and Research Bhopal,\\ Bhopal 462066, India}



\begin{abstract}
The strong gravitational field near massive blackhole is an interesting regime to test General Relativity(GR) and modified gravity theories. The knowledge of spacetime metric around a blackhole is a primary step for such tests. Solving field equations for rotating blackhole is extremely challenging task for the most modified gravity theories. Though the derivation of Kerr metric of GR is also demanding job, the magical Newmann-Janis algorithm does it without actually solving Einstein equation for rotating blackhole. Due to this notable success of Newmann-Janis algorithm in the case of Kerr metric, it has been being used to obtain rotating blackhole solution in modified gravity theories. In this work, we derive the spacetime metric for the external region of a rotating blackhole in a nonlocal gravity theory using Newmann-Janis algorithm. We also derive metric for a slowly rotating blackhole by perturbatively solving field equations of the theory. We discuss the applicability of Newmann-Janis algorithm to nonlocal gravity by comparing slow rotation limit of the metric obtained through Newmann-Janis algorithm with slowly rotating solution of the field equation.

\end{abstract}

\maketitle
\section{INTRODUCTION}
Recent gravitational wave observations done by Advanced LIGO/VIRGO \cite{Abbott:2016blz,Abbott:2016nmj,Abbott:2017vtc,Abbott:2017gyy,Abbott:2017oio} and an image captured by Event Horizon Telescope\cite{Akiyama:2019cqa,Akiyama:2019eap} has rejuvenated interest about a mysterious object called `blackhole' in scientific community all over the world. Naively speaking, blackhole is a star collapsed under its own gravity to a point where the curvature of the spacetime diverges to infinity. The spacetime surrounding static blackhole is described by Schwarzschild metric. It is a static spherically symmetric vacuum solution of the Einstein equation. 
Due to its strong gravitational field, region near a blackhole can be a good laboratory for the tests of any gravity theory. Since most astrophysical objects rotate, the blackhole created from their gravitational collapse is more likely to be a rotating blackhole. The rotating blackhole solution of the Einstein equation was given by R.P.Kerr in 1963\cite{Kerr:1963ud}. It is a stationary, axially symmetric vacuum solution. The Kerr blackhole displays many interesting properties which have been studied over the years\cite{Geroch:1970cd,Carter:1971zc,Hansen:1974zz,Bekenstein:1973ur,Wald:1999vt,Fiola:1994ir,Nikolic:2009ju,Bradler:2013gqa,Susskind:1997ci,Hawking:1971ei,Carr:1974nx,Carr:1975qj,Meszaros:1975ef,Cha1975}. 

In 1965, Newmann and Janis shown that by means of complex coordinate transformations operated on Schwarzschild metric one obtains a new metric. They investigated the new metric and found that the new metric corresponds to a massive ring rotating about its axis of symmetry\cite{nj}. This method was also shown to be working between Reissner-Nordstr{\"o}m and Kerr-Newmann metric. The generalization of this method in the presence of cosmological constant was done by Demia\'{n}ski\cite{Demianski:1972uza}. Gradually it became very popular to derive the rotating blackhole metric because it avoids all mathematical difficulties involved in solving Einstein equation and is generally known as Newmann-Janis (NJ) or Demia\'{n}ski-Janis-Newman (DJN) algorithm.

Initial developments in Newmann-Janis algorithm were majorly restricted to General Relativity(GR). But, discovery of late-time cosmological acceleration\cite{Nobel2011} and quest for quantum gravity provided thrust to look for modified gravity theories beyond GR. In the absence of fundamental direction, these studies are being done mostly on trial and error basis. Various modified gravity theories are constructed and tested against astrophysical and cosmological observations. The detection of gravitational waves made strong field tests of modified gravity possible and therefore it is important to know the structure of the rotating blackhole metric for modified gravity theories. 

Lately, Newmann-Janis method has been used comfortably as a way to derive the Kerr metric in different modified gravity theories\cite{Yazadjie:2000,Kyriakopoulos:2009sf,Ghosh:2012ji,Cornell:2017irh,Kumar:2018pkb}. But, it should be done with caution since the metric generated from the NJ method may not be the solution of the field equations of a particular theory. In fact some cases of failure of NJ algorithm in non-GR theories have been reported in \cite{CiriloLombardo:2004qw,Hansen:2013owa}. In this work, we show that the metric generated from the NJ algorithm does not match with the metric derived by solving field equations in a nonlocal gravity model.

In recent studies Nonlocal gravity has emerged as an effective candidate for cosmological constant. The first nonlocal model was studied by Deser and Woodard in 2008 where they considered nonlocal correction given by $Rf\left( \frac{1}{\Box}\right)$\cite{Woodard:2014iga}. In this work we consider a specific model, called $RR$ model, that has correction term proportional to $R\frac{1}{{\Box}^{2}}R $ proposed in\cite{Maggiore:2014sia}. The complete action for $RR$ model is given by
\begin{equation}
S = \frac{1}{2\kappa^{2}}\int d^{4}x \sqrt{-g}\Big[ R + \frac{m^{2}}{3} R\frac{1}{\Box^2}R\Big] + \mathcal{L}_{m},  \label{action}
\end{equation}
where $m$ is the mass scale associated with nonlocal correction to the Einstein-Hilbert(EH) action. In the limit $m \rightarrow 0 $, the above action \eqref{action} reduces to EH action. This model is studied substantially in \cite{Calcagni:2010ab,Dirian:2014xoa,Nersisyan:2016hjh,Fernandes:2017vvo,Tian:2019bla}.
The field equations\cite{Maggiore:2014sia} corresponding to action~\eqref{action} are
\begin{equation}
\begin{split}
\kappa^{2} T_{\alpha \beta} & =  G_{\alpha \beta} - \frac{m^{2}}{3}\Big\{ 2\Big(G_{\alpha
 \beta} - \nabla_{\alpha}\nabla_{\beta} + g_{\alpha \beta} \Box \Big)S  + g_{\alpha \beta} \nabla^{\gamma}U \nabla_{\gamma}S   - \nabla_{(\alpha}U \nabla_{\beta)}S  -\frac{1}{2}g_{\alpha \beta} U^{2} \Big\};\\&
 U = - \frac{1}{\Box}R,\;\;\;S = - \frac{1}{\Box}U,\label{fieldeq}
 \end{split}
 \end{equation}
where $G_{\alpha \beta}$ is the Einstein tensor and $T_{\alpha \beta}$ is the energy-momentum tensor of the matter. The spherically symmetric and static vacuum solution of the above field equation has been obtained in \cite{Maggiore:2014sia} on which we will apply the NJ algorithm to find the Kerr-like metric for $RR$ model. 

This paper is organized as follows: In Sec.~\ref{sec:rotatingmetric}, we derive the rotating blackhole metric for the RR model of the nonlocal gravity using DJN algorithm. In Sec.~\ref{sec:srbs}, we solve the field equations of $RR$ model for slowly rotating blackhole by perturbing spherically symmetric static solution of the model. Comparision of metrics obtained through two approaches mentioned above are done in Sec.~\ref{ssec:comparing}. Finally, we conclude our work in Sec.~\ref{sec:conclusion}


\section{Rotating Blackhole in RR model of Nonlocal Gravity}\label{sec:rotatingmetric}


It is by now well known that NJ algorithm does some ``trick" which transforms Schwarzschild metric into Kerr metric. The ``trick" is complexifying coordinates and then performing complex coordinate transformations. But, we still do not know how complexifying coordinates gives rotation to a static blackhole and exactly generate stationary axisymmetric vacuum solution of Einstein equation without actually solving the equation. The only requirement is the spherically symmetric static vacuum solution of the Einstein equation which works as a ``seed" metric. One can apply NJ method on the ``seed" metric and get rotating blackhole metric. Some studies have been done on conditions on properties of the ``seed" metric\cite{Yazadjie:2000,CiriloLombardo:2004qw}. 

The most discomforting aspect of the NJ algorithm is that there involves some arbitrariness in how one complexify the different functions of coordinates in the ``seed" metric\cite{Hansen:2013owa,Drake:1998gf,Bambi:2013ufa}. It lacks the solid base of physical argument on which one can choose how to complexify the coordinates. In the original NJ algorithm the choice was such that it gives Kerr metric in the end. This prevents the generalization of the NJ method especially when the rotating metric is not known apriori. Some efforts have been made to reduce the intrinsic arbitrariness of the method but still it remains elusive\cite{Drake:1998gf,Newman:2002mk}. Here, we apply the NJ algorithm on the spherically symmetric static solution of the field equations of the RR model of nonlocal gravity as written in \eqref{fieldeq} and obtain ``some" metric. We will compare this metric generated from NJ algorithm with solution of the field equations \eqref{fieldeq} in slow rotation limit in the next section.

Let us first write the seed metric which is Schwarzschild solution of the RR model, derived in \cite{Maggiore:2014sia}.
\begin{equation}
ds^{2} = -A(r) dt^2 + B(r) dr^{2} + r^{2}( d\theta^{2} + \sin^{2}\theta \;d\phi^{2} ), \label{2}
\end{equation}
where A(r) and B(r) are given by
\begin{equation}
A(r) \simeq 1 - \frac{2GM}{r}\Bigg(1 + \frac{m^{2}r^{2}}{6}\Bigg)  \label{3}
\end{equation}
\begin{equation}
B(r) \simeq A(r)^{-1} \label{4}
\end{equation}
The above metric is written in Boyer-Lindquist coordinates $(t, r, \theta, \phi)$. From cosmological point of view, the motivation of the RR model is to explain the late-time acceleration of the universe. Therefore the mass $m$ of the field corresponding to the correction term (the second term in the parenthesis in \eqref{3}) is of the order of $H_0$ where $H_0$ is the present value of Hubble parameter. Since the distance $r$ from the source is much smaller than the Hubble length $H_0^{-1}$, the authors of \cite{Maggiore:2014sia} derived Schwarzschild solution in region $r<<m^{-1}$ taking low-m expansion in the field equations. 

Here, we would like to make a note that the Ricci-flat solution is not the solution in our theory in case of spherical symmetry. As shown in \cite{Li:2015bqa,Briscese:2018bny,Briscese:2019rii}, any equation of motion(eom) which involves only Ricci scalar and its derivatives will be satisfied by the Ricci-flat metric since all terms vanish as $R$ vanishes. But, in our case the eom written in Eq.\eqref{fieldeq} does not admit Ricci-flat metric as its solution mainly because of terms like $\Box^{-1}R$ and $\Box^{-2}R$. To reach such conclusion, we have used arguments as follows. As discussed in \cite{Kehagias:2014sda}, the definition of $\Box^{-1}$ operator is such that the solution of the equation $\Box U = 0$ is $U=-\ln\left(1-\frac{2GM}{r}\right)$. If we substitute this solution of U in $\Box S = -U$ and solve it and then plug both the solutions of $U$ and $S$ in the field equation \eqref{fieldeq} then we can show that $R=0$ does not satisfy our field equation \eqref{fieldeq}.

The first step of the NJ algorithm is to transform the metric in \eqref{2} into Eddington-Finkelstein coordinates $(u, r, \theta, \phi)$ via
\begin{eqnarray}
dt = du + \Bigg(1 - \frac{2GM}{r}\Big(1 + \frac{m^{2}r^{2}}{6}\Big)\Bigg)^{-1} dr \label{6}
\end{eqnarray}
The metric \eqref{2} now can be written as
\begin{equation}
ds^{2} = -\left[1 - \frac{2GM}{r}\Bigg(1 + \frac{m^{2}r^{2}}{6}\Bigg)\right] \;du^2 - 2\;du\; dr +r^{2}\;( d\theta^{2} + \sin^{2}\theta \; d\phi^{2} ).\label{7} 
\end{equation}
Expressing the metric in \eqref{7} in terms of  null tetrad vectors as 
\begin{eqnarray}
g^{\mu\nu} = l^{\mu}n^{\nu} + l^{\nu}n^{\mu} - m^{\mu} \overline{m}^{\nu} - m^{\nu} \overline{m}^{\mu},   \label{8}
\end{eqnarray}
where the null tetrad vectors $ l^{\mu} , n^{\mu} , m^{\mu}$ and $ \overline{m}^{\mu}$ take the following form
\begin{eqnarray}
l^{\mu} &=& \delta_{1}^{\mu} \label{9} \\
n^{\mu} &=& \delta_{0}^{\mu} - \frac{1}{2} A(r) \delta_{1}^{\mu} \label{10} \\
m^{\mu} &=& \frac{1}{\sqrt{2}r}\Big( \delta_{2}^{\mu} + \frac{i}{\sin \theta} \delta_{3}^{\mu} \Big) \label{11} \\
\overline{m}^{\mu} &=& \frac{1}{\sqrt{2}r}\Big( \delta_{2}^{\mu} - \frac{i}{\sin \theta} \delta_{3}^{\mu} \Big). \label{12}
\end{eqnarray}
Now we complexify null tetrad vectors as
\begin{eqnarray}
l^{\mu} &=& \delta_{1}^{\mu} \label{9A} \\
n^{\mu} &=& \delta_{0}^{\mu} - \frac{1}{2} \tilde{A}(r) \delta_{1}^{\mu} \label{10A} \\
m^{\mu} &=& \frac{1}{\sqrt{2}\tilde{r}}\Big( \delta_{2}^{\mu} + \frac{i}{\sin \theta} \delta_{3}^{\mu} \Big) \label{11A} \\
\overline{m}^{\mu} &=& \frac{1}{\sqrt{2}r}\Big( \delta_{2}^{\mu} - \frac{i}{\sin \theta} \delta_{3}^{\mu} \Big), \label{12A}
\end{eqnarray}
where 
\begin{eqnarray}
\tilde{A}(r) &= & 1 - GM\left(\frac{1}{r} + \frac{1}{\tilde{r}}\right)\Big( 1 + \frac{1}{6} m^{2}r\tilde{r} \Big). \label{12B}
\end{eqnarray}
Here, $\tilde r$ is the complex conjugate of $r$. One can note that complexifying of $1/r$ and $r^2$ are done differently. Apply complex transformation 
\begin{equation}
{x'}^{\rho} = x^{\rho} + ia \cos \theta (\delta^{\rho}_{0} - \delta^{\rho}_{1} ).
\end{equation}
After this step tetrad vector becomes 
\begin{eqnarray}
l'^{\mu} &=& \delta_{1}^{\mu} \label{13} \\
n'^{\mu} &=& \delta_{0}^{\mu} -\frac{1}{2} \tilde{A}(r',\theta) \delta_{1}^{\mu} \label{14}
\end{eqnarray}
\begin{equation}
m'^{\mu} = \frac{1}{\sqrt{2}(r' + i a \cos \theta)}\left( i a \sin \theta ( \delta _{0}^{\mu}-\delta_{1}^{\mu}) + \delta_{2}^{\mu}  + \frac{i}{\sin \theta} \delta_{3}^{\mu} \right) \label{15} 
\end{equation}
\begin{equation}
\overline{m'}^{\mu} = \frac{1}{\sqrt{2}(r' - i a \cos \theta)}\left( -i a \sin \theta ( \delta _{0}^{\mu}-\delta_{1}^{\mu}) + \delta_{2}^{\mu} - \frac{i}{\sin \theta} \delta_{3}^{\mu} \right),\label{16} 
\end{equation}
where, 
\begin{eqnarray}
\tilde{A}(r',\theta) &= & 1 - \frac{2GMr'}{\Sigma}\Big( 1 + \frac{1}{6} m^{2}\Sigma \Big) \label{17} \\
\Sigma & = & r'^{2} + a^{2}\cos ^{2}\theta.\label{18}
\end{eqnarray}
Using Eq.\eqref{8}, one can read off the components of the contravaraint metric from tetrad vectors found in the above step as
\begin{eqnarray}
g^{\mu\nu}=
  \begin{bmatrix}
    -\frac{a^{2} \sin^{2}\theta}{\Sigma} & 1 + \frac{a^{2}\sin^{2}\theta}{\Sigma} & 0 & -\frac{a}{\Sigma}  \\
   1 + \frac{a^{2}\sin^{2}\theta}{\Sigma}  & -\tilde{A} -\frac{a^{2}\sin^{2}\theta}{\Sigma} & 0 & \frac{a}{\Sigma}  \\
    0 & 0 & -\frac{1}{\Sigma} & 0  \\
    -\frac{a}{\Sigma} & \frac{a}{\Sigma} & 0 & -\frac{1}{\Sigma \sin^{2}\theta}  \label{19}
  \end{bmatrix}
\end{eqnarray}
Here we have replaced $r'$ by $r$. Inverting metric \eqref{19} to get the covariant metric 
\small
\begin{eqnarray}
g_{\mu\nu}=
  \begin{bmatrix}
    -\tilde{A} & -1 & 0 &- a \sin^{2} \theta (1 -\tilde{A})  \\
   -1  & 0 & 0 & a \sin^{2}\theta \\
    0 & 0 & \Sigma & 0  \\
    - a \sin^{2} \theta (1 -\tilde{A}) & a \sin^{2}\theta & 0 &  \sin^{2}\theta \Big( \Sigma + a^{2} \sin^{2} \theta (2 - \tilde{A})\Big) \label{20}
  \end{bmatrix}
\end{eqnarray}
\normalsize
To convert the metric (\ref{20}) into Boyer-Lindquist coordinates, we perform following transformations
\begin{eqnarray}
du &=& dt' -\frac{r^2+a^2}{\Delta} dr, \label{21} \\
d\phi &=& d\phi' - \frac{a}{\Delta} dr, \label{22}
\end{eqnarray}
where we define 
\begin{equation}
\Delta = \Sigma \tilde{A}(r,\theta) + a^{2}\sin^{2}\theta.  \label{23} 
\end{equation}
Thus output line element of the spacetime can be written as 
\begin{equation}
ds^{2}=   -\tilde{A}\; dt^{2} - 2a\sin^{2}\theta\left[1 - \tilde{A}\right]\;dt\;d\phi + \dfrac{\Sigma}{\Delta}\;dr^{2}  + \Sigma \;d\theta^{2} + {\rm sin}^{2}\theta\left[\Sigma + (2 -\tilde{A}) a^2 \rm sin^{2}\theta \right]\;d\phi^{2}.
\label{25}
\end{equation}
The above metric is written in the form like $g_{\mu\nu}^K + b_{\mu\nu}$, where $g_{\mu\nu}^K$ is the Kerr metric of GR and $b_{\mu\nu}$ is the correction terms due to modified gravity except $g_{rr}$ component. One can also express the $g_{rr}$ component in the same way. Finally rewriting the metric derived in \eqref{25} as
\begin{eqnarray}
ds^2 = -\left[1-\frac{2GMr}{\Sigma}-\frac{2GMrm^2}{6}\right]\;dt^2 - \left[\frac{2GMr}{\Sigma}2a\sin^2\theta + \frac{2GMrm^2}{6}2a\sin^2\theta\right]\;dt\;d\phi + \left[\frac{\Sigma}{\Delta_{GR}} +\right.\nonumber\\
\left.\frac{2GMr\Sigma^2m^2}{6\Delta^2_{GR}}\right] dr^2 + \Sigma d\theta^2 + \left[\sin^2\theta\left\{\Sigma + \left(1+\frac{2GMr}{\Sigma}\right) a^2\sin^2\theta\right\} +\sin^2\theta\left\{\frac{2GMrm^2a^2\sin^2\theta}{6}\right\}\right]\;d\phi^2,    
\label{26}    
\end{eqnarray}
where $\Delta_{GR} \equiv r^2 +a^2 - 2GMr$.


\section{Slowly Rotating Blackhole Solution of RR Model}\label{sec:srbs}
\begin{figure}[!ht]
  \centering
 \includegraphics[scale=0.7]{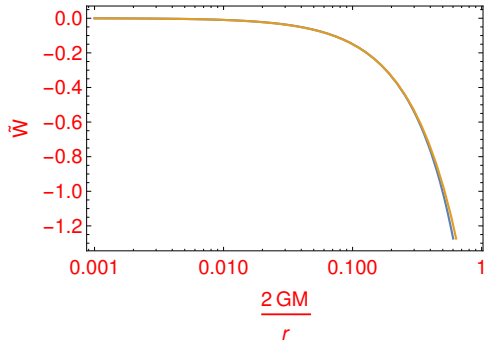}
\caption{$\widetilde{W}$ Vs. $2GM/r$. Blue and yellow curves show numerical solution of Eq.\eqref{41} and analytical solution of Eq.\eqref{41b} for $\widetilde{W}$ respectively.}
\label{WtildeVst} 
\end{figure}

In this section, we derive the metric for the slowly rotating blackhole for the nonlocal model given in \eqref{action} and compare it with the slow rotation limit of the metric derived in \eqref{25}. We consider Schwarzschild metric which is spherically symmetric and static solution of the Einstein equation as background metric, 
\begin{equation}
    ds^2 = -\left[1-\frac{2GM}{r}\right]\;dt^2 + \left[1-\frac{2GM}{r}\right]^{-1}\;dr^2 + r^2\;(d\theta^2 + \sin^{2}\theta\;d\phi^2).\label{a27}
\end{equation}
Now, first order perturbations are the terms having either first order in $m^2$ or first order in spin $a$. The terms having combined order of $m^2a$ will be considered as second order terms along with $a^2$ order terms. As we switch on the first order term $m^2$ our governing equation will be no more Einstein equation instead it will be Eq.\eqref{fieldeq} and coefficients of $dt^2$ and $dr^2$ will be replaced by $A(r)$ and $B(r)$ given in equations \eqref{3} and \eqref{4}. Then, up to first order in $m^2$, the exterior region of a spherically symmetric and static blackhole can be described by the metric
\begin{equation}
    ds^2 = -A(r)\;dt^2 + B(r)\;dr^2 + r^2(\;d\theta^2 + \sin^{2}\theta\;d\phi^2),\;\;where\;  \nonumber 
\end{equation}
\begin{equation}
 A(r) = 1-\frac{2GM}{r}-\frac{GMm^2r}{3}\;and\; B(r) = A(r)^{-1}.\label{27}
\end{equation}
 We can verify that setting $m^{2}= 0$ in Eqns. \eqref{3} and \eqref{4} the standard general relativistic forms of $A(r)$ and $B(r)$ are recovered. 
Now let us switch on the second order perturbation which will include terms of the order $a^2$ and $m^2a$. Then the metric \eqref{27} becomes
\begin{eqnarray}
ds^2= -\left[A(r)+\frac{2a^2GM}{r^3}\cos^{2}\theta\right]\;dt^2 + \left[B(r)+\frac{a^2}{r^2}\left(1-\frac{2GM}{r}\right)^{-1}\left(\cos^{2}\theta-\left(1-\frac{2GM}{r}\right)^{-1}\right)\right]\;dr^2  \nonumber\\
+\;[r^2 + a^2 \cos^{2}\theta]\;d\theta^2 + \left[r^2 \sin^{2}\theta + a^2\left(1+\frac{2GM}{r}\sin^{2}\theta\right)\right] d\phi^2 - r^{2}\sin^{2}\theta\; a\;w(r)\;d\phi\; dt.\label{33}
\end{eqnarray}
We introduce a function $w(r)$ as a co-factor of spin $a$ in $g_{t\phi}$ component. Since our desired metric is stationary it does not depend on $t$. The reason that the first order perturbation, i.e. of the order of $a$, only enters in $t\phi$ component of the metric is as follows. The time reversal symmetry and symmetry in the direction of spinning dictates that the only $t\phi$ component can have odd powers of $a$ \cite{Hartle:1967he}. From the Kerr metric in GR we know form of all the terms of the order of $a^2$. From Eq. \eqref{fieldeq}, one can calculate the $t\phi$ component of the field equation as
\begin{equation}
R_{t\phi}\left(1 - 2\frac{m^2}{3}S\right) = \frac{m^2}{3} \left[ g_{t\phi} U - 2\nabla_t\nabla_{\phi}S\right].\label{34}
\end{equation}
Writing the above equation up to first order in $m^2$ we get
\begin{equation}
    R_{t\phi} = \frac{m^2}{3} \left[ g_{t\phi} U - 2\nabla_t\nabla_{\phi}S\right].\label{34a}
\end{equation}
The solution of the field equations \eqref{fieldeq} give the expression $U=-\ln\left(1-\frac{2GM}{r}\right)$\cite{Kehagias:2014sda}. Since our system is independent of $t$ and $\phi$ the second term inside the square bracket in Eq.\eqref{34a} will vanish. Calculating $R_{t\phi}$ for the metric written in \eqref{33} and substituting it and other quantities in \eqref{34a} one can obtain differential equation for $w(r)$ as
\begin{eqnarray}
w'' + \frac{4}{r}w'+\frac{w}{r}\left[2\frac{A'}{A}-\frac{2}{r}\frac{1}{A}+\frac{2}{r}-\frac{2}{3}m^2\frac{r}{A}\ln\left(1-\frac{2GM}{r}\right)\right]=0. \label{35a}
\end{eqnarray}
Here, prime denotes the derivative with respect to $r$. After substituting the form of $A(r)$ and $A'(r)$ we obtain
\begin{eqnarray}
w'' + \frac{4}{r}w'-\frac{2}{3}m^2w\left[\frac{2GM}{r}+\ln\left(1-\frac{2GM}{r}\right)\right]\left(1-\frac{2GM}{r}\right)^{-1}=0. \label{35b}
\end{eqnarray}
For $m=0$, above equation reduces to its general relativistic counterpart\cite{Hartle:1967he} 
\begin{equation}
w'' + \frac{4}{r}w'= 0,\label{36}
\end{equation}
whose solution is given by $w_{GR}=4GM/r^3$. We can simplify \eqref{35b} if we absorb $r^2$ inside $w(r)$ in the metric \eqref{33} by defining $W(r)=r^2w(r)$. The differential equation \eqref{35b} can be written in terms of $W(r)$ as
\begin{eqnarray}
W''+W\left[-\frac{2}{r^2}-\frac{2}{3}m^2\left(\frac{2GM}{r} + \ln\left(1-\frac{2GM}{r}\right)\right)\left(1-\frac{2GM}{r}\right)^{-1}\right]=0.
\label{38}
\end{eqnarray}
We can make above differential equation dimensionless by defining $t\equiv 2GM/r$. Writing $A(r)$ in terms of $t$ as 
\begin{equation}
    A(t)=1-t-\frac{C^2}{6t},
    \label{39}
\end{equation}
where $C=2GMm$ which can be considered as a new coupling constant. The fact that the Schwarzschild radius $r_s(=2GM)<<m^{-1}$ implies $C^2<<1$. Writing \eqref{38} in terms of $t$ as
\begin{eqnarray}
    t^4\frac{d^2W}{dt^2}+2t^3\frac{dW}{dt}+W\left[-2t^2-\frac{2}{3}C^2\left(t+\ln(1-t)\right)(1-t)^{-1}\right]=0.
    \label{40}
\end{eqnarray}
The solution of \eqref{40} can be split into two parts as $W(t)=W_{GR}(t)+C^2\widetilde{W}(t)$, where $W_{GR}=r^2w_{GR}$ is the solution of \eqref{40} when $C=0$. We can remove the GR part from the \eqref{40} and end up with equation
\begin{equation}
    t^4\frac{d^2\widetilde{W}}{dt^2}+2t^3\frac{d\widetilde{W}}{dt}-2t^2\widetilde{W}-\frac{4}{3}t\left[t+\ln(1-t)\right](1-t)^{-1}=0,
    \label{41}
\end{equation}
up to order $C^2$. It is extremely difficult to solve the above equation analytically. To get an approximate analytic form of $\widetilde{W}(t)$ we first solve it numerically then do curve fitting with the numerical solution. We use mathematica for this. For numerical calculation, we consider initial conditions at radial distance far away from the blackhole where $r>>2GM$. In this limit, the term in multiplication with $C^2$ in Eq.\eqref{40} can be neglected and the contribution of $\widetilde{W}$ in $W(t)$ will be zero. Therefore, our initial conditions are $\widetilde{W}(0.0001)=0$ and $\widetilde{W}'(0.0001)=0$ at $t=0.0001$. 
For $r>>2GM$, the Eq.\eqref{41} can be approximated as
\begin{equation}
   t^4\frac{d^2\widetilde{W}}{dt^2}+2t^3\frac{d\widetilde{W}}{dt}-2t^2\widetilde{W}+\frac{4}{3}t\left(\frac{t^2}{2}\right)(1+t)=0.
    \label{41b} 
\end{equation}
The analytical solution of above equation is given by
\begin{eqnarray}
\widetilde{W}(t)= \alpha\;t^{-2}+ \beta\; t - \frac{1}{6}\;t^2- \frac{2}{9}\;t\; \ln(t).   
    \label{41c}
\end{eqnarray}
The constants $\alpha$ and $\beta$ are fixed by initial conditions and they come out to be $\alpha\approx 0$ and $\beta\approx -1.99625$. For clear visibility, we have shown a plot comprising numerical solution of Eq.\eqref{41} and analytical solution \eqref{41c} of Eq.\eqref{41b} in Fig.[\ref{WtildeVst}] and it can be seen that they match considerably well except the region near horizon. The fitted solution for $\widetilde{W}$ of Eq.\eqref{41} is given in the appendix \ref{appendix}. 
 If we consider the  solution written in \eqref{41c} is a good approximation then the metric for the exterior spacetime of a slowly rotating blackhole in case of $RR$ model is given by
\begin{eqnarray}
ds^2= -\left[1-\frac{2GM}{r}-\frac{2GMrm^2}{6}+\frac{2a^2GM}{r^3}\cos^{2}\theta\right]dt^2 +  \left[\frac{1}{1-\frac{2GM}{r}} + \frac{2GMrm^2}{6\left(1-\frac{2GM}{r}\right)^2}+\frac{a^2}{r^2}\left(1-\frac{2GM}{r}\right)^{-1}\right.\nonumber\\
\left.\left(\cos^{2}\theta-\left(1-\frac{2GM}{r}\right)^{-1}\right)\right]\;dr^2+ \;\left[r^2+ a^2 \cos^{2}\theta\right]\;d\theta^2 +\left[r^2 \sin ^{2}\theta+ a^2\left(1+\frac{2GM}{r}\sin^{2}\theta\right)\right]\;  d\phi^2\nonumber \\
- a\;\sin^{2}\theta\left[\frac{4GM}{r}+C^2\left( - 1.99625\left(\frac{2GM}{r}\right)- \frac{1}{6}\left(\frac{2GM}{r}\right)^2- \frac{2}{9}\left(\frac{2GM}{r}\right)  \ln\left(\frac{2GM}{r}\right)\right)\right]d\phi\; dt.\label{41a}
 \end{eqnarray}
 

\subsection{Comparing Two Metrics}\label{ssec:comparing}
Now our intention is to compare the two metrics (written in \eqref{26} and \eqref{41a}) to check the validity of NJ algorithm for $RR$ model. In order to do that we have to write down the slow rotation limit of the metric in \eqref{26}. Therefore, ignoring terms of the order higher than $a^2$, the metric \eqref{26} can be rewritten as 
\begin{eqnarray}
ds^2 = -\left[1-\frac{2GM}{r}-\frac{2GMrm^2}{6}+\frac{2a^2GM}{r^3}\cos^{2}\theta\right]\;dt^2 - \left[\frac{4GM}{r} + \frac{4GMrm^2}{6}\right]\;a\sin^2\theta\;dt\;d\phi+ \left[r^2+ a^2 \cos^{2}\theta\right] \nonumber \\
d\theta^2 \;+ \left[\frac{1}{1-\frac{2GM}{r}} + \frac{2GMrm^2}{6\left(1-\frac{2GM}{r}\right)^2}+\left(1-\frac{2GM}{r}\right)^{-1}\frac{a^2}{r^2}\left(\cos^{2}\theta-\left(1-\frac{2GM}{r}\right)^{-1}\right)\right]dr^2\;+ \nonumber\\
 \left[r^2 \sin ^{2}\theta+ a^2\left(1+\frac{2GM}{r}\sin^{2}\theta\right)\right]\;d\phi^2.\label{42}    
\end{eqnarray}
From Eqns.\eqref{41a} and \eqref{42}, it is clearly visible that the two metrics do not coincide with each other. It is noteworthy that the same analysis done in linearized gravity limit shows that the metric generated from NJ algorithm and one derived by solving field equations match with each other\cite{Kumar:2018pkb}.


\section{Conclusions}\label{sec:conclusion}
To conclude we have derived the spacetime metric for the exterior region of a rotating blackhole in a nonlocal gravity model called RR model. Furthermore, we have shown that the slow rotation limit of the metric generated by NJ algorithm applied to a spherically symmetric static solution of the model does not agree with the slowly rotating blackhole solution obtained by solving field equations themselves.

It is well known that there is an ambiguity about the NJ algorithm because one can carry out complexification step in Eq.\eqref{12B} in many different ways\cite{Drake:1998gf,Bambi:2013ufa}. We want to draw the attention to the point : the complexification in Eq.\eqref{12B} is done in such a way so that the output metric will reduce to (i) Kerr metric in $m\rightarrow 0$ limit and (ii) seed metric of Eq.\eqref{2} in $a\rightarrow 0$ limit. The nonlocal part of the function $ A(r)$ is unchanged after the complexification as seen in Eq.\eqref{17}. We have also checked that even if we do complexification of $1/r$ and $r^2$ in Eq.\eqref{12B} in different way we will end up with the output metric such that, when expanded in terms of rotation parameter $a$, it will be same as the metric written in Eq.\eqref{42} up to second order. There may be difference in higher order terms which we neglect for slow rotation limit. We are not sure that a different complexification will reduce the discrepancy between two slowly rotating metric \eqref{41a} and \eqref{42} for terms having higher order than second order.

The two metrics differ by power of coordinate $r$ only in $g_{t\phi}$ component. The rotated metric via NJ method involves $g_{t\phi}$ term having positive power of $r$ while slowly rotating solution has negative power series in terms of $r$. This can result into a drastically divergent physical scenario. Thus, our investigation gives rise to suspicion about the applicability of NJ algorithm to modified gravity theories or at least in case of present model of nonlocal gravity.  

The question that what the expected properties of the rotating blackhole metric for it to be a physically/astrophysically a viable object should be, is outside the scope of this work. A thorough study in this direction and on possible modification of NJ algorithm can resolve the issue and present a trustworthy method to derive the Kerr-like solutions in modified gravity theories.


\section{Acknowledgement} This work was partially supported by DST grant number SERB/PHY/2017041. This work was carried out at IISER Bhopal. We would like to thank the EPJC referee for his insightful comments.

\appendix
\section{Fitted Form of $\widetilde{W}$}\label{appendix}
\begin{eqnarray}
    \widetilde{W}^{fit}(t) = 0.00367496 - 1.27855\; t - 2.44258\;t^2 -3.51342\;t^3 + 49.1209\;t^4 - 140.155\;t^5 + 115.164 \; t^6 \nonumber\\
 +91.2408\;t^7 - 82.8183\;t^8 -129.952\;t^9 - 25.6561\;t^{10} 
 + 99.2167\; t^{11} + 132.56\;t^{12} + 57.3422\;t^{13} - 63.8041\;t^{14}   \nonumber\\
 -142.463\; t^{15} - 188.1\;t^{16} +6.80019\;t^{17} + 152.833\;t^{18}  +166.905\;t^{19} - 164.073\; t^{20},
    \label{app1}
\end{eqnarray}
where $t=\frac{2GM}{r}$.



\end{document}